\documentclass[11pt,1p]{elsarticle}
\newcommand{\ignore}[1]{}
\usepackage{a4wide}
\usepackage[T1]{fontenc}
\usepackage{lmodern}

\ignore{
\documentclass[5p]{elsarticle}
\usepackage[fontsize=9.5]{scrextend}
\usepackage{fullpage}
\usepackage[uppercase,charter]{mathdesign}
}

\makeatletter
\def\ps@pprintTitle{%
 \let\@oddhead\@empty
 \let\@evenhead\@empty
 \def\@oddfoot{\centerline{\thepage}}%
 \let\@evenfoot\@oddfoot}
\makeatother

\usepackage{balance}
\usepackage{enumitem}
\setitemize{noitemsep,topsep=2pt,parsep=0pt,partopsep=0pt}
\setenumerate{noitemsep,topsep=2pt,parsep=0pt,partopsep=0pt}
\usepackage{amsmath}
\usepackage{amsthm} 
\usepackage{amsfonts}
\usepackage{xcolor}
\usepackage[colorlinks]{hyperref}
\hypersetup{colorlinks=true,linkcolor=gruen,citecolor=gruen}
\definecolor{gruen}{rgb}{0,.5,0}

\usepackage{hyperref}
\usepackage[capitalise,nameinlink]{cleveref}
\crefformat{equation}{(#2#1#3)}
\crefname{appendix}{}{}
\crefname{lem}{Lemma}{Lemmas}
\crefname{clm}{Claim}{Claims}
\crefname{thm}{Theorem}{Theorems}

\newtheorem{thm}{Theorem}
\newtheorem{lem}{Lemma}

\theoremstyle{remark}
\newtheorem{rem}{Remark}

\newcommand{\f}{\mathcal{F}}
\newcommand{\RR}{\mathbb{R}}
\newcommand{\pRR}{\mathbb{R}_{+}} 
\newcommand{\mst}[2]{\mathrm{mst}(#2)} 
\newcommand{\mmst}[2]{\mathrm{mst}_{#1}(#2)}
\newcommand{\dist}[2]{\mathrm{dist}_{#2}(#1)} 
\newcommand{\ee}{f} 
\newcommand{\TT}{T^*} 

\newcommand{\exchange}[3]{{#1}\!-\!{#2}\!+\!{#3}}
\newcommand{\xx}{x'} 
\newcommand{\xxx}{y} 
\newcommand{\FW}[2]{D_{#1}^{#2}} 

\begin{document}
\begin{frontmatter}
\title{Sorting Can Exponentially Speed Up Pure Dynamic Programming}

\author[1]{Stasys~Jukna\fnref{fn1}}
\ead{stjukna@gmail.com}
\fntext[fn1]{Research supported by the DFG grant
    JU~3105/1-2 (German Research Foundation).}

\author[2]{Hannes Seiwert\corref{cor1}}
\cortext[cor1]{Corresponding author.}
\ead{seiwert@thi.cs.uni-frankfurt.de}

\address[1]{Department of Mathematics and Computer Science,
Vilnius University, Lithuania}
\address[2]{Institute of Computer Science, Goethe University, Frankfurt am Main, Germany}

\begin{abstract}
Many discrete minimization problems, including various versions of
  the shortest path problem, can be efficiently solved by dynamic
  programming (DP) algorithms that are ``pure'' in that they only perform basic
  operations, as $\min$, $\max$, $+$,
  but no conditional branchings via if-then-else in their recursion equations. It is known that any pure $(\min,+)$ DP algorithm solving the minimum weight spanning tree problem on undirected $n$-vertex graphs must
  perform at least $2^{\Omega(\sqrt{n})}$ operations.
We show that this problem \emph{can} be solved by a pure  $(\min,\max,+)$ DP algorithm performing only $O(n^3)$
operations. The algorithm is essentially a $(\min,\max)$ algorithm: addition operations are only used to output the final values. The presence of both $\min$ and $\max$ operations means that
now DP algorithms can sort: this explains the title of the paper.
\end{abstract}

\end{frontmatter}

\section{Introduction}
A discrete $0$-$1$ optimization problem is specified by giving a
finite set $E$ of \emph{ground elements} together with a family $\f\subseteq 2^E$ of subsets
of these elements, called \emph{feasible solutions}. The problem
itself is, given an assignment of nonnegative real weights to the
ground elements, to compute the minimum or the maximum weight of a
feasible solution, the latter being the sum of weights of its elements. Note that we only need $(\min,+)$ or $(\max,+)$ operations to \emph{define} such problems.

For example, in the \emph{assignment problem} $E$ is the set of all edges of a complete bipartite graph and $\f$ is the family of all perfect matchings in it, each viewed as set of its edges. In the \emph{MST problem} (minimum weight spanning tree problem) on a connected graph $G=(V,E)$, feasible solutions are spanning trees of $G$, etc.

Dynamic programming (DP) is a fundamental algorithmic paradigm for
solving such optimization problems. Many DP algorithms are
\emph{pure} in that they only perform basic operations,  as $\min$, $\max$, $+$, $-$, in their recursion equations, but no \emph{conditional branchings} via if-then-else or argmin/argmax, or other additional operations. In particular, the recursions then do not
depend on the actual input weightings.

Notable examples of pure DP algorithms are the
Bell\-man--Ford--Moore algorithm for the shortest $s\text{-}t$ path
problem \cite{bellman,ford,Moore1957}, the Floyd--Warshall
algorithm for the all-pairs shortest paths problem
\cite{floyd,warshall}, the Held--Karp DP algorithm for the traveling
salesman problem \cite{held62} and the Dreyfus--Levin--Wagner
algorithm for the weighted Steiner tree problem~\cite{dreyfus,levin}.
The Viterbi $(\max,\times)$ DP
algorithm~\cite{viterbi} is also a pure $(\min,+)$ DP algorithm via
the isomorphism $h:(0,1]\to\RR_+$ given by $h(x)=-\ln x$.

There are, however, important optimization problems that can
be efficiently solved using greedy-type algorithms, but cannot be
efficiently solved by pure $(\min,+)$ or $(\max,+)$ DP algorithms. One of such problems, is the famous
\emph{MST problem} on an undirected connected graph $G=(V,E)$, which we have already mentioned above: given an assignment
$x:E\to\pRR$ of nonnegative real weights to the edges of $G$, compute the minimum weight
$\mst{G}{x}=\mmst{G}{x}$ of a spanning tree of~$G$:
\[
\mst{G}{x}=\min\{x(T)\colon \mbox{$T$ is a spanning tree of $G$}\}\,,
\]
where $x(T)=\sum_{e\in T}x(e)$; here and throughout $\pRR$ stands for
the set of all nonnegative real numbers.
In the \emph{directed} version of the MST problem, known as the
\emph{minimum arborescence problem}, the underlying graph $G$ is
directed and one seeks for the minimum weight of an arborescence
of~$G$; an \emph{arborescence} of a digraph $G$ is a directed tree in
which all vertices of $G$ are reachable by directed paths from one
fixed root vertex.

That every pure $(\min,+)$ DP algorithm for the  minimum arborescence problem on the complete $n$-vertex graph $G=K_n$  must perform  $2^{\Omega(n)}$ operations was proved by Jerrum and Snir in their seminal paper~\cite{jerrum}.
As we have recently shown in~\cite{JS19}, even the
simpler \emph{undirected} MST problem requires $2^{\Omega(\sqrt{n})}$ operations.
So, pure $(\min,+)$ DP algorithms for both these problems must perform an \emph{exponential} in $n$ number of operations.

Therefore, the following result of Fomin, Grigoriev and
Koshevoy~\cite{FGK} came as a surprise.
Using ideas from the electrical engineering (Kirchhoff's effective conductance formula and the star-mesh transformation to compute effective conductances), they show that
both (the directed and the undirected) MST problems \emph{can} be solved by pure $(\min,+,-)$ DP algorithms performing only $O(n^3)$ operations.
That is,
\begin{itemize}
\item \emph{subtraction} can exponentially speed up pure $(\min,+)$ DP algorithms.
\end{itemize}

In this paper, we show that, in fact, the MST problem can already be solved by a pure $(\min,\max,+)$ DP algorithm performing only $O(n^3)$ operations (\cref{thm:MST} below).
 Hence, already the \emph{monotone} $\max$ operation, instead of the \emph{non-monotone} subtraction $(-)$  operation, can exponentially speed up pure $(\min,+)$ DP algorithms.
The presence of \emph{both} $\min$ and $\max$ operations means that
now DP algorithms can sort; this explains the title of this paper:
\begin{itemize}
\item already \emph{sorting} can exponentially speed up pure $(\min,+)$ DP algorithms.
\end{itemize}
Note that $(\min,+,-)$ operations can be (easily) simulated by $(\min,\max,+)$ operations because $\max(x,y)=-\min(-x,-y)$, but not vice versa.

\section{Our results}
Let $G=(V,E)$ be an undirected connected graph.
Given a weighting $x:E\to\pRR$ of the edges of $G$, the \emph{min-max
  distance} between two vertices $u$ and $v$, which we denote by $\dist{u,v}{x}$, is the minimum, over all
paths from $u$ to $v$ in $G$, of the maximum weight of an edge along
this path:
\[
\dist{u,v}{x}=\min_P\ \max\{x(\ee)\colon \ee\in P\}\,,
\]
where the minimum is taken over all paths $P$ in $G$ between the vertices $u$ and $v$.
That is, the min-max distance between vertices $u$ and $v$
is the minimum number $d$ for which there is a path in $G$ between $u$
and $v$ with all edges of weight at most~$d$. The min-max distance $\dist{e}{x}$ of
an edge $e=\{u,v\}$ is the min-max distance between its endpoints $u$
and $v$. Note that the min-max distance of any edge does
not exceed its weight (the edge itself is a path between its
endpoints), but may be smaller, that is, we always have $\dist{e}{x}\leq x(e)$.

The following theorem relates min-max distances to the MST problem.
\begin{thm}\label{thm:main}
  Let $G=(V,E)$ be an undirected $n$-vertex graph, and
  $T=\{e_1,\ldots,e_{n-1}\}$ be a spanning tree of $G$.  Then
  for every weighting $x:E\to\pRR$, we have
   \[
  \mst{G}{x}=\dist{e_1}{x_0\!} + \dist{e_2}{x_1\!} + \cdots +
  \dist{e_{n-1}}{x_{n-2\!}}\,,
  \]
  where $x_0=x$, and each next weighting $x_i:E\to\pRR$ is obtained
  from $x$ by setting the weights of edges $e_1,\ldots,e_i$ to zero.
\end{thm}

\Cref{thm:main} allows us to efficiently solve the MST problem by a pure DP algorithm performing only $\min$, $\max$ and $+$ operations.
Namely, we can fix  an \emph{arbitrary} spanning tree $T$ of $G$; this tree $T$ will be used for \emph{all} arriving weightings $x:E\to\pRR$ of the edges of $G$. When an input weighting $x$ arrives, compute the min-max distances of the $n-1$ edges of
  the (fixed) tree $T$ under the corresponding modifications of the weighting $x$ by the Floyd--Warshall DP algorithm. By \cref{thm:main}, the sum of all these distances
  is then exactly the minimum weight of any spanning tree of $G$ with
  respect to the input weighting $x$. This yields a pure $(\min,\max,+)$ DP algorithm performing $O(n^4)$ operations. Some additional savings (see \cref{sec:algo} for details) lead to the following theorem.

\begin{thm}\label{thm:MST}
  The MST problem on every undirected connected graph on $n$ vertices
  can be solved by a pure $(\min,\max,+)$ DP algorithm performing $O(n^3)$ operations.
\end{thm}

\begin{rem}\label{rem:malpani}
Hu~\cite{hu} reduced
the problem of computing all
  min-max distances to the MST problem.
  When an input weighting $x$ of the
  edges arrives, find a spanning tree $T_x$ of $G$ of minimal
  $x$-weight. Then, with respect to this weighting, the min-max distance between any pair of
  vertices of $G$ is the maximal weight of an edge along the (unique)
  path in the tree $T_x$ between these vertices. That is, all min-max distances in the graph $G$ and in the minimum spanning $T$ are \emph{identical}.
  This result was re-discovered (with a more detailed proof) by Malpani and Chen~\cite[Theorem~2.1]{malpani}.

  Our \cref{thm:main} does the \emph{converse} reduction: it reduces
  the MST problem to the min-max distance problem.
\end{rem}

\begin{rem}
That the MST problem is related to the min-max distances was observed already by Maggs and Plotkin~\cite{plotkin}. They consider the case when weights of edges are distinct; hence, for every such weighting $x:E\to\pRR$, the minimum weight spanning tree $T_x$ is unique. They show that then $T_x=\{e\in E\colon \dist{e}{x}=x(e)\}$. This  result also gives a $(\min,\max,+)$ DP algorithm for the MST problem: use the Floyd--Warshall DP algorithm to compute the min-max distances $\dist{e}{x}$ of all edges $e$, and then sum up the weights of all edges for which $\dist{e}{x}=x(e)$ holds.

The main difference of this algorithm  from that given by \cref{thm:MST}  (besides the restriction to distinct weights, which is not crucial) is that it essentially uses \emph{conditional branchings}: {\sf if}  $\dist{e}{x}=x(e)$ {\sf then} accept $e$ {\sf else} reject $e$. Thus, the DP algorithm in~\cite{plotkin} is \emph{not} a pure DP algorithm.
In contrast, our algorithm uses no conditional branchings: it just performs $(\min,\max)$ operations to compute the min-max distances of $n-1$ edges (of one, fixed in advance, spanning tree), and then just uses $+$ operations to output the sum of these values. Thus, \cref{thm:MST} removes the need of conditional branchings in the Maggs--Plotkin DP algorithm, and does this without increasing the total number of performed operations.
\end{rem}

\section{Proof of Theorem~\ref{thm:main}}
Since each next weighting in \cref{thm:main} sets the weight of one single edge to zero, it is enough to consider what happens after each such setting.

\begin{lem}\label{lem:main}
Let $G=(V,E)$ be an undirected connected graph. Then for every weighting $x:E
\to \pRR$, and for every edge $e\in E$, we have
\begin{equation}\label{eq:matroids}
  \mst{G}{x}=\mst{G}{\xx}+\dist{e}{x}\,,
\end{equation}
where $\xx:E\to\pRR$ is the weighting obtained from $x$ by giving zero weight to the
edge $e$, and leaving other weights unchanged.
\end{lem}

\Cref{lem:main} immediately yields \cref{thm:main}
because after the weights of all edges $e_1,\ldots,e_{n-1}$ of the tree $T$
are set to zero, we have an optimal spanning tree  $T$ of \emph{zero} weight,
that is, $\mst{G}{x_{n-1}}=x_{n-1}(T)=0$; recall that all weights are
\emph{nonnegative}.

\begin{proof}[Proof of \cref{lem:main}]
We prove \cref{eq:matroids} by showing the inequalities
\begin{equation}\label{eq:1}
\mst{G}{\xx}\leq\mst{G}{x}-\dist{e}{x}
\end{equation}
and
\begin{equation}\label{eq:2}
\mst{G}{x}\leq\mst{G}{\xx}+\dist{e}{x}
\end{equation}
separately. To show \cref{eq:1}, let $T$ be a spanning tree of $G$ of minimal $x$-weight. If $e\in T$, then $\xx(T)=x(T)-x(e)$. Since $x(e) \geq \dist{e}{x}$ and $x(T)=\mst{G}{x}$,  inequality \cref{eq:1} trivially holds in this case.

Assume now that $e\not\in T$. We claim that there is an edge $\ee\in T$ of weight $x(\ee) \geq \dist{e}{x}$ such that $\TT =\exchange{T}{\ee}{e}$ is a spanning tree of $G$.
To show this, take the (unique) path $P$ in
  the tree $T$ between the endpoints of $e$. Let $\ee\in T$ be an edge
  of that path of maximal weight $x(\ee)$. By the definition of
  $\dist{e}{x}$, \emph{every} path between the endpoints of $e$ must
  contain an edge of $x$-weight at least $\dist{e}{x}$. Hence,
  $x(\ee)\geq \dist{e}{x}$.
The removal of the edge $\ee$ from $T$ cuts the tree $T$ into two connected components. Since the set \mbox{$P+e$} forms a cycle, the edge $e$ lies between these two components. Thus, $\TT=\exchange{T}{\ee}{e}$ is a spanning tree of $G$, and  inequality \cref{eq:1} follows:
 \begin{align*}
\mst{G}{\xx} &\leq \xx(\TT) = \xx(T)-\xx(\ee) + \xx(e)\\
  &=x(T)-x(\ee)
   \leq x(T)-\dist{e}{x} \\
  &= \mst{G}{x} -\dist{e}{x}\,.
\end{align*}

To show \cref{eq:2}, we use the fact that the $\xx$-weight
$\xx(e)=0$ of the edge $e$ is the \emph{smallest} possible weight
(all weights are nonnegative). So, $e\in T$ holds for at least one spanning tree $T$ of $G$ of minimal $\xx$-weight; fix such a tree~$T$.

We claim that  there is
an edge $\ee$ of $G$ of weight $x(\ee) \leq \dist{e}{x}$ such that $\TT =\exchange{T}{e}{\ee}$ is a spanning tree of $G$. Indeed, by the definition of $ \dist{e}{x}$, there is
a path $P$ in $G$ between the endpoints of $e$
  such that $x(\ee)\leq \dist{e}{x}$ holds for all edges $\ee\in P$. The path $P$ does not need to lie in the tree $T$, but at
  least one edge $\ee\in P$ must cross the cut induced by the edge $e$
  of $T$, that is, must lie between the two connected components of $T$ after the edge $e$ is removed. Thus, $\TT=\exchange{T}{e}{\ee}$ is also a spanning tree of $G$.

So, since $\xx(T)=x(T-e)$ holds, inequality \cref{eq:2} follows:

\begin{align*}
  \mst{G}{x} &\leq x(\TT) = x(T-e) + x(\ee)
   =\xx(T)+x(\ee)\\
   &\leq \xx(T) + \dist{e}{x}
   = \mst{G}{\xx} + \dist{e}{x}\,. \!\!\!\!
  \qedhere
\end{align*}
\end{proof}

\section{Proof of Theorem~\ref{thm:MST}} \label{sec:algo}

Let $G=(V,E)$ be an undirected connected graph with $V=\{1,2,\ldots,n\}$. Our goal is to show that
the MST problem on $G$ can
  be solved by a pure DP algorithm performing $O(n^3)$ $(\min,\max,+)$ operations.

  First, we can easily reduce the MST problem on $G$ to the MST
  problem on the complete graph $K_n$ on~$V$. For an input
  weighting $x:E\to\pRR$, compute the maximum weight
  $M=\max\{x(e)\colon e\in E\}$ with $|E|-1=O(n^2)$ $\max$
  operations. Then give the weight $M$ to every non-edge of
  $G$. Under the resulting weighting $\xxx:K_n\to\pRR$, we have
  $\mst{G}{x}=\mst{K_n}{\xxx}$. What we achieved is that now
  \emph{all} pairs of distinct vertices, not only
  the edges of~$G$, are weighted edges.

  Now, given a weighting $x:K_n\to\pRR$, the \emph{max-length} of a
  walk is the weight of its heaviest edge. Hence, the min-max distance
  $\dist{e}{x}$ of an edge $e=\{i,j\}$ is the minimal max-length of a
  walk between $i$ and $j$. Note that this minimum will always be
  achieved on some simple \emph{path} between $i$ and~$j$: every walk
  between $i$ and $j$ contains a path between $i$ and $j$.  The
  min-max distances $\dist{e}{x}$ of \emph{all} edges $e$ of $K_n$ can be
  simultaneously computed by the Floyd--Warshall DP algorithm~\cite{floyd,warshall} as follows.

  A $k$-\emph{walk} is a walk using only vertices from
  $\{1,\ldots,k\}$ as inner vertices.  As subproblems, we take
  $\FW{i,j}{k}$ = the minimum max-length over all $k$-walks $P$
  between vertices $i$ and $j$.  Initial values are the weights
  $\FW{i,j}{0}=x(i,j)$ of the edges $\{i,j\}$ of $K_n$. Every $k$-walk
  between $i$ and $j$ either does not go through the vertex $k$, or
  does. So, the recurrence is:
  \[
  \FW{i,j}{k}=\min\left\{\FW{i,j}{k-1},\
    \max\{\FW{i,k}{k-1}, \FW{k,j}{k-1}\}\right\}
  \]
  Then $\FW{i,j}{n}=\dist{i,j}{x}$ is the min-max distance between $i$
  and~$j$. Hence, all min-max distances $\dist{i,j}{x}$ can be simultaneously
  computed with $N=O(n^3)$ $\min$ and $\max$ operations.

  According to \cref{thm:main}, we only have to compute
  min-max distances $\dist{e_1}{x_0}, \dots$,  $\dist{e_{n-1}}{x_{n-2}}$ of $n-1$ edges $e_1, \dots, e_{n-1}$ (of a fixed spanning tree $T$), and add them together. This gives us a pure
  DP algorithm solving the MST problem on any $n$-vertex graph by
  performing $O(nN+n-1)=O(n^4)$ $(\min,\max,+)$ operations.

  But, since in our case each next weighting differs from the previous one on only one edge, we can reduce the total number of operations to $O(n^3)$. Compute all min-max distances $\dist{i,j}{x}$ under the initial
  weighting $x$ using the Floyd--Warshall algorithm, as above. After
  that, it is enough just to update these weights. Namely, the next to
  $x$ weighting $\xx$ only sets the weight of one edge
  $e=\{a,b\}$ to $0$, and leaves the
  weights of other edges unchanged.

  \balance

  Every path from a vertex $i$ to a vertex $j$ either goes through the
  edge $e$, or not. If a path of minimal $\xx$-max-length does not
  go through $e$, then $\dist{i,j}{\xx}=\dist{i,j}{x}$. If a path of minimal $\xx$-max-length goes through $e$, then $\dist{i,j}{\xx}$ is the minimum of
  $\max(\dist{i,a}{x},\dist{b,j}{x} )$ and
    $\max( \dist{i,b}{x}, \dist{a,j}{x})$,
  because the edge $e\!=\!\{a,b\}$ can be entered from both its
  endpoints. Thus, $\dist{i,j}{\xx}$ is the minimum of $\dist{i,j}{x}$
  and  $\max(\dist{i,a}{x},\dist{b,j}{x} )$ and $\max(
  \dist{i,b}{x}, \dist{a,j}{x})$.

  We thus can compute the min-max distances
  between all pairs of vertices under the next to $x$ weighting $\xx$
  performing only $K=O(n^2)$ additional $(\min,\max)$ operations. Since we only have to update the distances $n-2$ times, the total number of performed operations is $N+(n-2)K+n-1=O(n^3)$.
 \qed


\begin{thebibliography}{10}

\bibitem{bellman}
R.~Bellman.
\newblock On a routing problem.
\newblock {\em Quarterly of Appl. Math.}, 16:87--90, 1958.

\bibitem{dreyfus}
S.E. Dreyfus and R.A. Wagner.
\newblock The {S}teiner problem in graphs.
\newblock {\em Networks}, 1(3):195--207, 1971.

\bibitem{floyd}
R.W. Floyd.
\newblock Algorithm 97, shortest path.
\newblock {\em Comm. ACM}, 5:345, 1962.

\bibitem{FGK}
S.~Fomin, D.~Grigoriev, and G.~Koshevoy.
\newblock Subtraction-free complexity, cluster transformations, and spanning
  trees.
\newblock {\em Found. Comput. Math.}, 15:1--31, 2016.

\bibitem{ford}
L.R. Ford.
\newblock Network flow theory.
\newblock Technical Report P-923, The Rand Corp., 1956.

\bibitem{held62}
M.~Held and R.M. Karp.
\newblock A dynamic programming approach to sequencing problems.
\newblock {\em SIAM J. on Appl. Math.}, 10:196--210, 1962.

\bibitem{hu}
T.C. Hu.
\newblock The maximum capacity route problem.
\newblock {\em Oper. Res.}, 9:898--900, 1961.

\bibitem{jerrum}
M.~Jerrum and M.~Snir.
\newblock Some exact complexity results for straight-line computations over
  semirings.
\newblock {\em J. ACM}, 29(3):874--897, 1982.

\bibitem{JS19}
S.~Jukna and H.~Seiwert.
\newblock Greedy can beat pure dynamic programming.
\newblock {\em Inf. Process. Lett.}, 142:90--95, 2019.

\bibitem{levin}
A.Y. Levin.
\newblock Algorithm for the shortest connection of a group of graph vertices.
\newblock {\em Sov. Math. Dokl.}, 12:1477--1481, 1971.

\bibitem{plotkin}
B.M. Maggs and S.A. Plotkin.
\newblock Minimum-cost spanning tree as a path-finding problem.
\newblock {\em Inf. Process. Lett.}, 26(6):291--293, 1988.

\bibitem{malpani}
N.~Malpani and J.~Chen.
\newblock A note on practical construction of maximum bandwidth paths.
\newblock {\em Inf. Process. Lett.}, 83(3):175--180, 2002.

\bibitem{Moore1957}
E.F. Moore.
\newblock The shortest path through a maze.
\newblock In {\em Proc. Internat. Sympos. Switching Theory}, volume~II, pages
  285--292, 1957.

\bibitem{viterbi}
A.~Viterbi.
\newblock Error bounds for convolutional codes and an asymptotically optimum
  decoding algorithm.
\newblock {\em IEEE Trans. on Information Theory}, 13(2):260--269, 1967.

\bibitem{warshall}
S.~Warshall.
\newblock A theorem on boolean matrices.
\newblock {\em J. ACM}, 9:11--12, 1962.

\end{thebibliography}
\end{document}